\documentclass[aps,prl,reprint,amsmath,amssymb,nofootinbib,floatfix]{revtex4-2}
\usepackage{graphicx} 
\usepackage{physics}
\usepackage{letltxmacro}
\usepackage{hyperref}
\usepackage{verbatim}

\usepackage{ulem}

\renewcommand{\Re}{\real}
\renewcommand{\Im}{\imaginary}
\providecommand{\ii}{\text{i}}

\usepackage[dvipsnames]{xcolor}

\newcommand{\bisp}{\boldsymbol{\psi}}
\newcommand{\inst}[1]{\boldsymbol{\mathcal{#1}}}

\newcommand{\unit}[1]{\mathbf{\hat{#1}}}
\LetLtxMacro{\originaleqref}{\eqref}
\renewcommand{\eqref}{Eq.~\originaleqref}

\date{\today}
\usepackage{cleveref}

\begin{document}

\title{Electromagnetic symmetry dislocations}

\author{Alex J. Vernon}
\email{alexander.j.vernon@kcl.ac.uk}
\affiliation{Department of Physics, King's College London, Strand, London WC2R 2LS, UK}
\affiliation{London Centre for Nanotechnology}

\author{Sebastian Golat}
\affiliation{Department of Physics, King's College London, Strand, London WC2R 2LS, UK}
\affiliation{London Centre for Nanotechnology}

\author{Francisco~J. Rodr\'iguez-Fortu\~no}
\email{francisco.rodriguez\_fortuno@kcl.ac.uk}
\affiliation{Department of Physics, King's College London, Strand, London WC2R 2LS, UK}
\affiliation{London Centre for Nanotechnology}

\begin{abstract}
Singular optics aims to understand and manipulate light's topological defects, pioneered by the discovery that phase vortex lines, strands of destructive interference, naturally occur in scalar wave fields.
Monochromatic electromagnetic fields, however, are described by complex three-dimensional vectors that make individual scalar phase vortices in their vector components, which depend on the choice of co-ordinate basis, less meaningful.
Instead, polarisation singularities can capture the vector texture of complicated, even non-paraxial light, with separate spatial descriptions for the electric $\mathbf{E}$ and magnetic $\mathbf{H}$ fields.
But polarisation textures, too, are basis-dependent, because the laws of electromagnetism can be expressed not only by separate $\mathbf{E}$ and $\mathbf{H}$ fields, but by linear combinations of the two.
We instead propose fundamental, basis-independent topological features generic in monochromatic electromagnetic fields: one- and two-dimensional structures that relate to time-averaged symmetries, including parity, duality and time-reversal, held locally by the combined electric and magnetic field polarisation geometry.

\end{abstract}

\maketitle

\paragraph*{Introduction}\!\!\!\!---\!\,\,\,%
Scalar dislocation lines (or phase vortices) have long been identified as general phenomena in scalar wave superpositions \cite{Nye1974,DennisThesis}, wherein perfect destructive interference of the complex scalar field $\phi$, the total field of the superposition, defines stable one-dimensional line structures in three real-space dimensions.
Satisfying two real conditions $\Re\{\phi\}=\Im\{\phi\}=0$, which can informally be associated with the singularity's codimension (2), a dislocation line disturbs the surrounding field to the extent that the field's phase angle, $\arg\phi$, smoothly realises all values between $0$ and $2\pi$ an integer number of times along any circuit containing the line in isolation.
A linearly polarised electromagnetic field with homogeneous direction, for instance, can be described by a single complex scalar field whose dislocation lines leave behind imprints in the streamlines of momentum density.
But light is a vector wave and in arbitrary superpositions of monochromatic plane waves, in near fields, or in focussed beams, both electric and magnetic field phasors are three-dimensional vectors spanning $\mathbb{C}^3$ and only in their individual scalar components do dislocation lines exist, by themselves unable to effect a mechanical consequence in the vector field as a whole. 
They depend on the choice of basis in $\mathbb{C}^3$, and in that sense are not fundamental; for instance, phase singularities of the electric field's complex Cartesian components $E_x\unit{x}+E_x\unit{y}+E_z\unit{z}$ are different to those in spherical components $E_r\unit{r}+E_\theta\boldsymbol{\hat{\theta}}+E_\phi\boldsymbol{\hat{\phi}}$.

The natural extension of phase singularities to complex vector fields like $\mathbf{E}$ appear when considering lines of polarisation \cite{Nye1983,Nye1987,Berry2001,Dennis2002,Angelsky2023,Barnett2023}, of which there are two important types: circular and linear polarisation, constructing C lines and L lines.
Polarisation is independent of whatever $\mathbb{C}^3$ basis (Cartesian, spherical, cylindrical, etc) is used to represent the electric or magnetic field, and C lines and L lines, corresponding to the two extremites of the generic polarisation ellipse that is swept out by a time-harmonic field vector, are called polarisation singularities.
Since many materials tend to break electric-magnetic symmetry, a description of light's topological skeleton using electric \cite{Garcia2017,Vernon2023} or magnetic \cite{Peng2022,Fu2024} polarisation singularities and other quantities is usually most relevant (and experimentally measurable) yet, particularly in the presence of matter breaking other symmetries or in free space, not strictly fundamental.
What is almost always neglected is that polarisation \textit{is} basis-dependent, but within an adjacent vector space.

The electromagnetic bispinor $\bisp$ \cite{Berry2009,Bliokh2014,Barnett2014,Dennis2023,Golat2024} (or `wavefunction') shows us how, combining the electric and magnetic fields in a separate, two-component complex vector space (the $\mathbb{C}^2$ space):
\begin{equation}\label{EM_bispinor}
    \bisp(\mathbf{r})=\frac{1}{2}\begin{pmatrix}
        \sqrt{\varepsilon_0}\mathbf{E}\\\sqrt{\mu_0}\mathbf{H}
    \end{pmatrix}=\begin{pmatrix}
        \mathbf{F}_\text{e}\\\mathbf{F}_\text{m}
    \end{pmatrix}.
\end{equation}
Phasors $\mathbf{E}$ and $\mathbf{H}$ can be thought of as complex vector-valued components of the $\mathbb{C}^2$ vector $\bisp$.
We could equivalently express \eqref{EM_bispinor} as $\bisp=\mathbf{F}_\text{e}\otimes\unit{e}+\mathbf{F}_\text{m}\otimes\unit{m}$, making explicit that it is represented in an electric-magnetic basis and with basis vectors $\unit{e}$ and $\unit{m}$ which, distinguishing the two rows of \eqref{EM_bispinor}, describe the orientation of $\bisp$ in $\mathbb{C}^2$.
The familiar $\mathbb{C}^3$ vector $\mathbf{E}$ we call the electric field, not by coincidence of notation, but specifically because it is associated with the unit vector $\unit{e}$ in this other vector space.

Naturally, the basis of the bispinor may be changed; its entries admit linear combinations of $\mathbf{E}$ and $\mathbf{H}$ \cite{Golat2024}.
For example, a basis of left- or right-handed pure-helicity fields $\bisp=\mathbf{F}_\text{R}\otimes\unit{R}+\mathbf{F}_\text{L}\otimes\unit{L}$ \cite{Golat2024,FernandezCorbaton2013,Aiello2015a,Berry2019} can be chosen,
\begin{equation}\label{LR_bispinor}
    \bisp(\mathbf{r})=\bisp_\text{RL}(\mathbf{r})=\pmqty{\mathbf{F}_\text{R}\\\mathbf{F}_\text{L}}=\frac{1}{2\sqrt{2}}\pmqty{\sqrt{\varepsilon_0}\mathbf{E}+i\sqrt{\mu_0}\mathbf{H}\\\sqrt{\varepsilon_0}\mathbf{E}-i\sqrt{\mu_0}\mathbf{H}},
\end{equation}
in which the basis vectors $\unit{R}=(\unit{e}-i\unit{m})/\sqrt{2}$ and $\unit{L}=(\unit{e}+i\unit{m})/\sqrt{2}$ correspond to the rows of $\bisp_\text{RL}$, or even a parallel and anti-parallel field pair $\bisp=\mathbf{F}_\text{p}\otimes\unit{p}+\mathbf{F}_\text{a}\otimes\unit{a}$,
\begin{equation}\label{PA_bispinor}
   \bisp(\mathbf{r})=\bisp_\text{PA}(\mathbf{r})=\pmqty{\mathbf{F}_\text{p}\\\mathbf{F}_\text{a}}=\frac{1}{2\sqrt{2}}\pmqty{\sqrt{\varepsilon_0}\mathbf{E}+\sqrt{\mu_0}\mathbf{H}\\\sqrt{\varepsilon_0}\mathbf{E}-\sqrt{\mu_0}\mathbf{H}},\\
\end{equation}
where $\unit{p}=(\unit{e}+\unit{m})/\sqrt{2}$ and $\unit{a}=(\unit{e}-\unit{m})/\sqrt{2}$.
In each representation prefactors have been chosen such that the inner product $\bisp^\dagger\cdot\bisp$ produces the time-averaged electromagnetic energy density.

While in the singular optics literature the polarisation topology of mainly the electric field, and sometimes magnetic field, are well understood, their study is from a mathematical standpoint an arbitrary choice and there is no reason not to devote attention to the polarisation structure of (for example) $\mathbf{F}_\text{R}$ or $\mathbf{F}_\text{a}$ which, too, contain polarisation singularities of their own that do not generally coincide with those of other $\mathbb{C}^3$ fields.
In the same way as the normal vectors to the $\mathbf{E}$ and $\mathbf{H}$ polarisation ellipses contribute to spin angular momentum density, geometric aspects of polarisation ellipses in other $\mathbb{C}^2$ bases correspond to other dynamic electromagnetic quantities (see supplementary information).

There is good reason, we suggest, to try to find singular structures that correspond to arguably more fundamental characteristics of light, and which do not depend on the choice of basis in $\mathbb{C}^2$.
That is our purpose here.
We identify six structures in total: three that relate directly to parity, time-reversal and duality symmetries held locally by the time-harmonic geometry of the electromagnetic field vectors, averaged over time, and a further three corresponding to compound symmetries.
Half of these structures are phase singularities in complex scalar fields, each codimension 2, defining stable one-dimensional strands permeating the electromagnetic field, while the other three are two-dimensional surfaces.
Five of them are to our knowledge unrecognised, while the remaining structure, a stationary, time-averaged version of the Riemann-Silberstein vortex \cite{BialynickiBirula_2003,Berry2004_2,Kaiser2004,Nye2019}, we do not believe is understood in relation to symmetries held by field vectors.

\paragraph*{Lines of time-averaged symmetry}\!\!\!\!---\!\,\,\,%
Each of our proposed structures are located where the electromagnetic field vectors are locally symmetric to the effect of one or a sequence of parity-inversion $\hat{P}$, time-reversal $\hat{T}$ or duality operations $\hat{D}$, when averaged over one cycle.
An operator $\hat{O}$ transforms in a general basis the bispinor $\bisp=(\mathbf{F}_i,\mathbf{F}_j)^T$ according to,
\begin{equation}\label{bisp_transformation}
    \hat{O}\bisp=\bisp'=\begin{pmatrix}
        \mathbf{F}_i'\\\mathbf{F}_j'
    \end{pmatrix},
\end{equation}
The action of $\hat{P}$, $\hat{T}$ and $\hat{D}$ on the components of the bispinor $\bisp$ depend on its chosen basis; in the electric-magnetic representation \eqref{EM_bispinor}, for example, the symmetry operators in $\mathbb{C}^2$ reflect or rotate the bispinor and possibly conjugate its components \cite{Bliokh2014}: 
\begin{equation}
\begin{split}\label{symmetries}
    \{\hat{P},\hat{T},\hat{D}\}&\bisp_\text{EM}\\&=\frac{1}{2}\left\{\begin{pmatrix}-\sqrt{\varepsilon_0}\mathbf{E}\\\sqrt{\mu_0}\mathbf{H}\end{pmatrix},
                                                            \begin{pmatrix}\sqrt{\varepsilon_0}\mathbf{E}^*\\-\sqrt{\mu_0}\mathbf{H}^*\end{pmatrix},
                                                            \begin{pmatrix}\sqrt{\mu_0}\mathbf{H}\\-\sqrt{\varepsilon_0}\mathbf{E}\end{pmatrix}\right\}.
\end{split}
\end{equation}
We define the electromagnetic field at a position $\mathbf{r}_\text{0}$ to be locally symmetric with respect to a certain transformation $\hat{O}$ if the transformed component coefficients of the bispinor, $\mathbf{F}_i'$ and $\mathbf{F}_j'$, can be superimposed with the original coefficients, pre-transformation, via a rotation $\hat{R}$:
\begin{equation}\label{symmetry_to_O}
    \begin{pmatrix}
        \hat{R} & 0\\0 & \hat{R}
    \end{pmatrix}\bisp'(\mathbf{r}_\text{s})=\bisp(\mathbf{r}_\text{s}).
\end{equation}
But it is important to specify in which space the complex phasors $\mathbf{F}_i'$ and $\mathbf{F}_j'$ are rotated.
If there exists a real rotation matrix $\hat{R}$ belonging to SO$(3)$ that can recover $\mathbf{F}_i$ and $\mathbf{F}_j$ from $\mathbf{F}_i'$ and $\mathbf{F}_j'$, then the electromagnetic field is symmetric with respect to the transformation $\bisp'=\hat{O}\bisp$ for all values of time.
This in the most generic non-paraxial fields only occurs under exceptional circumstances because SO$(3)$ is the group of rotations in real space, meaning that the four real vectors $\mathbf{p}_{i,j}$ and $\mathbf{q}_{i,j}$ where $\mathbf{F}_{i,j}=\mathbf{p}_{i,j}+\ii\mathbf{q}_{i,j}$ must all be eigenvectors of $\hat{R}$, with specific eigenvalues dictated by the operator $\hat{O}$.

If instead we designate $\hat{R}$ as an SU$(3)$ matrix, a rotation in a complex inner product space, then \eqref{symmetry_to_O} describes \textit{time-averaged} symmetry of the field vectors with respect to $\hat{O}$.
This results from the fact that scalar time-averaged quantities are complex inner products, which are by definition preserved by unitary matrices.
Time-averaged symmetry means that while the geometry of the two instantaneous vectors, $\inst{F}_{i,j}(\mathbf{r},t)=\Re\{\mathbf{F}_{i,j}\exp(-i\omega t)\}$, and the degree to which they break $\hat{O}$-symmetry changes from moment to moment, they oscillate between opposite symmetry-breaking configurations that cancel out over the course of one cycle.
This, too, places constraints on the field phasor geometry.
Nevertheless, these particular constraints are not overly severe and, depending on the operator in question, are often satisfied along one-dimensional strands that develop naturally in general non-paraxial light, much like polarisation singularities.

We have quite deliberately left the bispinor basis unspecified in this section, assigning it the non-specific vector-valued components $\mathbf{F}_i$ and $\mathbf{F}_j$.
For it is often possible for us to choose the bispinor basis so that $\mathbf{F}_i$ and $\mathbf{F}_j$ are eigenvectors of $\hat{O}$ with distinct eigenvalues, meaning,
\begin{equation}\label{O_transformation}
    \hat{O}\mathbf{F}_i=\lambda_1\mathbf{F}_i\quad\hat{O}\mathbf{F}_j=\lambda_2\mathbf{F}_j.
\end{equation}
Equation (\ref{symmetries}) has an example of this behaviour for $\hat{O}=\hat{P}$ in the electric-magnetic bispinor basis.
Symmetry with respect to $\hat{O}$ means that \eqref{O_transformation} can be replicated by the rotation matrix $\hat{R}$ in place of $\hat{O}$, or more explicitly, that $\mathbf{F}_i$ and $\mathbf{F}_j$ are also eigenvectors of $\hat{R}$ with the same eigenvalues $\lambda_{1,2}$, where $\lambda_1\neq\lambda_2$.
Since the three eigenvectors of an SU$(3)$ matrix form an orthonormal basis, if $\mathbf{F}_i$ and $\mathbf{F}_j$ are to be eigenvectors of $\hat{R}$ with different eigenvalues then $\expval{\mathbf{F}_i,\mathbf{F}_j}=\mathbf{F}_i^*\cdot\mathbf{F}_j=0$.

\begin{figure}[t]
    \centering
    \includegraphics[width=\columnwidth]{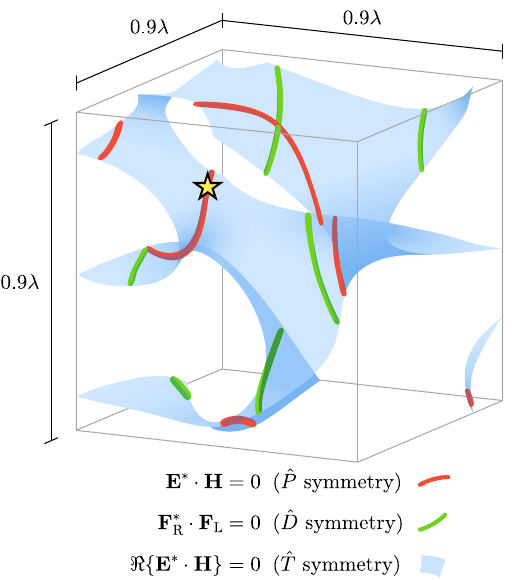}
    \caption{Cycle-averaged symmetry structures shown in a $(0.9\lambda)^3$ volume produced by interfering four monochromatic, randomly polarised plane waves with random wavevector directions.
    The electric and magnetic polarisation ellipses (that satisfy $\mathbf{E}^*\cdot\mathbf{H}=0$) at the location of the star are plotted in Fig. 2.
    }
    \label{figure1}
\end{figure}

Whatever the choice of bispinor basis, the quantity $\mathbf{F}_i^*\cdot\mathbf{F}_j$ is a complex scalar field.
Such is the nature of any complex scalar field in three dimensions, including the superposition of interfering scalar waves, that the zeros of $\mathbf{F}_i^*\cdot\mathbf{F}_j$ are one-dimensional structures forming just as naturally in non-paraxial fields as polarisation singularities.
The topological structures relating to time-averaged symmetry of the local field vector geometry with respect to certain fundamental symmetry operators behave in precisely this way, as shown in Fig. \ref{figure1}.
We define these structures next, beginning with $\hat{P}$.

\paragraph*{Parity, duality and time-reversal}\!\!\!\!---\!\,\,\,%
An object that cannot after parity inversion be superimposed with its original image is chiral.
Light can both adopt a chiral spatial structure, such as in the helical phase fronts of a vortex beam, and be chiral locally depending on field polarisation (individually the $\mathbf{E}$ or $\mathbf{H}$ field is not locally chiral because a polarisation ellipse is not itself a chiral shape \cite{Ayuso2019,Pisanty2019}, but the combined geometry of the two field vectors can be).
According to \eqref{symmetries}, the electric and magnetic field phasors are eigenvectors of the parity operator with eigenvalues $\pm1$,
\begin{equation}
    \hat{P}\mathbf{E}=-\mathbf{E}\quad\hat{P}\mathbf{H}=\mathbf{H},
\end{equation}
which suggests that the scalar dislocation line defined by $\mathbf{E}^*\cdot\mathbf{H}=0$ locates positions in an electromagnetic field where the field vectors are, cycle-averaged, locally parity-symmetric.
It is important to stress the word \textit{locally} here, because we do not consider the inversion of the whole spatial distribution of the field by transformation of the position vector $\mathbf{r}\rightarrow-\mathbf{r}$.
To instead specify the local parity of the field vectors at a given position $\mathbf{r}_0$, we designate $\mathbf{r}_0$ as the inversion centre (which is after all an arbitrary choice).

\begin{figure}[t]
    \centering
    \includegraphics[width=\columnwidth]{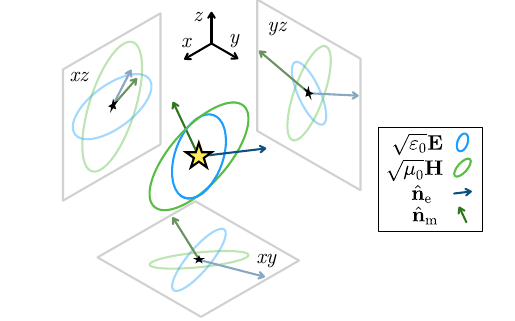}
    \caption{The electric (blue) and magnetic (green) polarisation ellipses and normal vectors at the location of the star in Fig. 1, along with projections of the ellipses onto the $xy$, $yz$ and $xz$ planes.
    Despite their non-distinct 3D orientation, these ellipses satisfy $\mathbf{E}^*\cdot\mathbf{H}=0$ and are cycle-averaged $\hat{P}$-symmetric.}
    \label{figure2}
\end{figure}

Because in paraxial light the real part of $\mathbf{E}^*\cdot\mathbf{H}$ is zero (among other vanishing quantities), it is tempting to connect the chirality of the electric and magnetic field vector geometry with rather simplistic features of electric and magnetic polarisation ellipses (paraxial $\mathbf{E}$ and $\mathbf{H}$ fields are related by a cross product with a well-defined wavevector, meaning their polarisation ellipses are identical up to a rotation by $90^\circ$ and lie in the same plane).
Towards the non-paraxial regime, however, it becomes extremely difficult to picture the specific relationship between electric and magnetic field vectors and their time evolution that means that at a certain location, they can be on average superimposed with their locally $\hat{P}$-inverted images.
For it is quite possible for both electric and magnetic fields to be elliptically polarised, and their polarisation ellipses oriented seemingly indistinctly relative to each other like in Fig. \ref{figure2}, and still satisfy $\mathbf{E}^*\cdot\mathbf{H}=0$.
Typically it is the helicity density $h=-\Im\{\mathbf{E}^*\cdot\mathbf{H}\}/(2\omega c)$ \cite{Cameron2012,Bliokh2014,Bliokh2013} that is associated with the local chirality of an electromagnetic field (confusingly, chirality density is also the name of a different quantity which is distinct from helicity in time-dependent fields \cite{Tang2010,Mackinnon2019}).
But $h$ is strictly speaking a measure of the difference in number density of left-handed and right-handed photons \cite{Trueba1996,Afanasiev1996,Golat2024}, and should not be misconstrued as a direct indicator of the chirality of the local electric and magnetic field vector geometry.
We argue that time-averaged local parity symmetry requires vanishing of both real and imaginary parts of $\mathbf{E}^*\cdot\mathbf{H}$, which correspond to $h$ as well as the reactive helicity density $h_\text{reac}=-\Re\{\mathbf{E}^*\cdot\mathbf{H}\}/(2\omega c)$ \cite{NietoVesperinas2021,Bliokh2014} being zero simultaneously.

Next, the duality operator $\hat{D}$, for which the vectors $\mathbf{F}_\text{R}$ and $\mathbf{F}_\text{L}$ from \eqref{LR_bispinor} are eigenvectors with eigenvalues $\pm i$,
\begin{equation}
    \hat{D}\mathbf{F}_\text{R}=i\mathbf{F}_\text{R}\quad\hat{D}\mathbf{F}_\text{L}=-i\mathbf{F}_\text{L},
\end{equation}
indicating that time-averaged duality symmetry of the field vectors occurs when $\mathbf{F}_\text{R}^*\cdot\mathbf{F}_\text{L}$ is zero.
Notably, the two real scalar conditions satisfied here, $\Re\{\mathbf{F}_\text{R}^*\cdot\mathbf{F}_\text{L}\}=\Im\{\mathbf{F}_\text{R}^*\cdot\mathbf{F}_\text{L}\}=0$, are time-averaged versions of the fundamental invariants that vanish on Riemann-Silberstein vortices \cite{Berry2004_2}.
Riemann-Silberstein vortices are strands of plane-wave-like, null electromagnetic field existing even in non-paraxial fields, which are typically defined instantaneously \cite{BialynickiBirula_2003,Kaiser2004} and that way move through space rapidly \cite{Nye2017}.
Cycle-averaging introduced by the definition $\mathbf{F}_\text{R}^*\cdot\mathbf{F}_\text{L}=0$ renders the vortex stationary, as has been recognised \cite{Berry2004_2}, and therefore more realistically measurable in principle (but no longer Lorentz-invariant). 
As longitudinal field components diminish, all paraxial light approaches $\mathbf{F}_\text{R}^*\cdot\mathbf{F}_\text{L}=0$.

Finally, the time-reversal operator $\hat{T}$.
Unlike $\hat{P}$ and $\hat{D}$, none of the $\mathbb{C}^3$ vectors from \eqref{EM_bispinor} to \eqref{PA_bispinor} are eigenvectors of $\hat{T}$ due to complex conjugation.
Nonetheless, the PA representation of $\bisp$ \eqref{PA_bispinor} presents an interesting scenario where
\begin{equation}\label{T_transformation}
    \hat{T}\mathbf{F}_\text{p}=\mathbf{F}_\text{a}^*\quad\hat{T}\mathbf{F}_\text{a}=\mathbf{F}_\text{p}^*.
\end{equation}
If the vector geometry is cycle-averaged, time-reversal-symmetric, then a unitary matrix $\hat{R}$ should be able to undo the above transformation from right to left, but because a rotation cannot change the length of a vector, it must be that $|\mathbf{F}_\text{p}|=|\mathbf{F}_\text{a}|$, or alternatively, that $|\mathbf{F}_\text{p}|^2-|\mathbf{F}_\text{a}|^2\propto\Re\{\mathbf{E}^*\cdot\mathbf{H}\}=0$.

At the same time, unitarity of $\hat{R}$ means the inner product of the vectors pre- and post-rotation must be preserved, $\expval{\mathbf{F}_\text{a}^*,\mathbf{F}_\text{p}^*}=\expval{\mathbf{F}_\text{p},\mathbf{F}_\text{a}}$, which in this case is trivially true.
That means the single real condition $|\mathbf{F}_\text{p}|^2-|\mathbf{F}_\text{a}|^2\propto\Re\{\mathbf{E}^*\cdot\mathbf{H}\}=0$ is sufficient to ensure that $\hat{R}$ exists, and therefore indicates cycle-averaged $\hat{T}$-symmetry of the field vectors.
Since $\Re\{\mathbf{E}^*\cdot\mathbf{H}\}=0$ is also included in the conditions $\mathbf{E}^*\cdot\mathbf{H}=0$ and $\mathbf{F}_\text{R}^*\cdot\mathbf{F}_\text{L}=0$, the lines of time-averaged $\hat{P}$ and $\hat{D}$ symmetry are contained in the surfaces of time-averaged $\hat{T}$ symmetry, as shown in Fig. \ref{figure1}.

\paragraph*{Additional symmetries}\!\!\!\!---\!\,\,\,%
It is natural now to ask what the zeros of the complex scalar $\mathbf{F}_\text{p}^*\cdot\mathbf{F}_\text{a}$ correspond to.
In fact the $\mathbb{C}^3$ vectors in the PA basis are eigenvectors of the combined $\hat{P}\hat{D}$ operator,
\begin{equation}
    \hat{P}\hat{D}\mathbf{F}_\text{p}=-\mathbf{F}_\text{p}\quad\hat{P}\hat{D}\mathbf{F}_\text{a}=\mathbf{F}_\text{a}.
\end{equation}
implying that $\mathbf{F}_\text{p}^*\cdot\mathbf{F}_\text{a}=0$ corresponds generically to lines of cycle-averaged $\hat{P}\hat{D}$-symmetry.

Meanwhile, time-averaged symmetry of the field vectors with respect to other compound operators, namely $\hat{P}\hat{T}$ and $\hat{D}\hat{T}$, can be identified by using the EM and RL representations of $\bisp$:
\begin{equation}\label{DT}
    \hat{D}\hat{T}\sqrt{\varepsilon_0}\mathbf{E}=-\sqrt{\mu_0}\mathbf{H}^*,\quad\hat{D}\hat{T}\sqrt{\mu_0}\mathbf{H}=-\sqrt{\varepsilon_0}\mathbf{E}^*,
\end{equation}
\begin{equation}\label{PT}
    \hat{P}\hat{T}\mathbf{F}_\text{R}=-\mathbf{F}_\text{L}^*,\quad\hat{P}\hat{T}\mathbf{F}_\text{L}=-\mathbf{F}_\text{R}^*.
\end{equation}
Neither $\mathbf{E}$ nor $\mathbf{H}$ are eigenvectors of $\hat{D}\hat{T}$, and likewise, $\mathbf{F}_\text{R}$ nor $\mathbf{F}_\text{L}$ are eigenvectors of $\hat{P}\hat{T}$.
But the behaviour of \eqref{DT} and \eqref{PT} is similar to that of \eqref{T_transformation}, and with the same arguments as in the text which follows (that $\hat{R}$ cannot change the length of the vector it transforms, and must preserve inner products), we can associate time-averaged $\hat{D}\hat{T}$-symmetry with $\varepsilon_0|\mathbf{E}|^2-\mu_0|\mathbf{H}|^2=0$ and $\hat{P}\hat{T}$-symmetry with $|\mathbf{F}_\text{R}|^2-|\mathbf{F}_\text{L}|^2\propto\Im\{\mathbf{E}^*\cdot\mathbf{H}\}=0$.
These real scalar conditions define different two-dimensional surfaces corresponding to cycle-averaged $\hat{P}\hat{T}$-symmetry and $\hat{D}\hat{T}$-symmetry.

Even number combinations of $\hat{P}$, $\hat{D}$ and $\hat{T}$ may commute or anticommute, for example $\hat{P}\hat{D}\bisp=-\hat{D}\hat{P}\bisp$, though the conditions for time-averaged symmetry with respect to either permutation of the same two operators in a combination is the same.
Remarkably, expanding the expression $\mathbf{F}_\text{p}^*\cdot\mathbf{F}_\text{a}=0$ using the definitions of $\mathbf{F}_\text{p}$ and $\mathbf{F}_\text{a}$ in \eqref{PA_bispinor}, we see that $\varepsilon_0|\mathbf{E}|^2-\mu_0|\mathbf{H}|^2=0$ and $|\mathbf{F}_\text{R}|^2-|\mathbf{F}_\text{L}|^2\propto\Im\{\mathbf{E}^*\cdot\mathbf{H}\}=0$ are also satisfied, meaning the dislocation line $\mathbf{F}_\text{p}^*\cdot\mathbf{F}_\text{a}=0$ is also symmetric with respect to any even number combination of $\hat{P}$, $\hat{D}$ and $\hat{T}$ operators, averaged over each oscillation.
A summary of the time-averaged symmetry structures we propose and their generic dimension---the main results of this work---is given in Tab. \ref{table1}.
\begin{table}[h]\label{sym_table}
    \centering
    \begin{tabular}{c|c|c}
         Cycle-av. symmetry & Scalar field zero & Dim. \\\hline
         $\hat{P}$ & $\mathbf{E}^*\cdot\mathbf{H}=0$ & $1$\\\hline
         $\hat{D}$ & $\mathbf{F}_\text{R}^*\cdot\mathbf{F}_\text{L}=0$ & $1$\\\hline
         $\hat{T}$ & $|\mathbf{F}_\text{p}|^2-|\mathbf{F}_\text{a}|^2=0$ & $2$\\\hline
         $\hat{P}\hat{D}$ & $\mathbf{F}_\text{p}^*\cdot\mathbf{F}_\text{a}=0$ & $1$\\\hline
         $\hat{P}\hat{T}$ &  $|\mathbf{F}_\text{R}|^2-|\mathbf{F}_\text{L}|^2=0$ & $2$\\\hline
         $\hat{D}\hat{T}$ & $\varepsilon_0|\mathbf{E}|^2-\mu_0|\mathbf{H}|^2=0$ & $2$\\\hline
    \end{tabular}
    \caption{Structures in electromagnetic fields which correspond to time-averaged symmetries (first column) of the local field vector geometry, defined by zeros of complex or real scalar fields (second column) where $\mathbf{F}_\text{R/L}=(\sqrt{\varepsilon_0}\mathbf{E}\pm i\sqrt{\mu_0}\mathbf{H})/(2\sqrt{2})$ and $\mathbf{F}_\text{p/a}=(\sqrt{\varepsilon_0}\mathbf{E}\pm\sqrt{\mu_0}\mathbf{H})/(2\sqrt{2})$.
    The generic dimension of these structures in non-paraxial fields is in column three.}
    \label{table1}
\end{table}

\paragraph*{Conclusions}\!\!\!\!---\!\,\,\,%
We identified six fundamental topological structures permeating electromagnetic fields, on which certain symmetries or combinations of symmetries among parity inversion $\hat{P}$, duality $\hat{D}$ and time-reversal $\hat{T}$ are preserved by the local field vector geometry when averaged over one oscillation.
Of the six structures summarised in Tab. \ref{table1}, three are phase vortices (zeros in complex scalar fields) amounting to lines in real space, while the other three are defined by real scalar zeros, corresponding to surfaces in real space.
All structures map to axial lines or planes within the recently proposed ``electromagnetic symmetry sphere", a Bloch sphere that characterises the distribution of total energy density among (in appropriate units) helicity density, reactive helicity density and reactive energy density \cite{Golat2024}. 

We derived the conditions in Tab. \ref{table1} in the context of the most general non-paraxial fields, including near fields, focussed beams and random plane wave interference, where each of our proposed phase vortex lines and two-dimensional surfaces form naturally and stably.
In the specific case of far-field light the mathematical conditions of Tab. \ref{table1} are the same, but manifest differently in real space because of the restricted relationship between the fields and the well-defined wavevector in the paraxial regime.
Time-averaged $\hat{D}$, $\hat{T}$ and $\hat{D}\hat{T}$ symmetry is satisfied everywhere in paraxial far fields, while the phase vortices of $\mathbf{F}^*_\text{p}\cdot\mathbf{F}_\text{a}$ and $\mathbf{E}^*\cdot\mathbf{H}$ and the positions of local $\hat{P}\hat{T}$-symmetry coincide with lines of linear electric and magnetic polarisation.

Phase and polarisation singularities can be sculpted in such a way as to create exceptional, possibly skyrmionic topologies \cite{Dennis2010,Larocque2018,Sugic2021,Shen2022,Spaegele2023}.
We expect that the local symmetry structures in this work could be manipulated in much the same way.
The ability to locate strands and surfaces in electromagnetic fields corresponding to certain local symmetries of the field vectors could also be useful for observing special light-matter interactions of interest, including optical forces arising from certain broken symmetries of matter \cite{Bliokh2014,Golat2024_2}.
A similar methodology to that which we have used could be applied to other wave fields \cite{MuelasHurtado2022,Smirnova2024,Kille2024}, including linearised gravity \cite{DennisThesis,Golat2020}, to uncover (depending on the degrees of freedom of the wave) analogous local symmetry structures.


\bibliography{bibliography}

\end{document}


\title{Supplementary Information for `Electromagnetic symmetry dislocations'}



\author{Alex J. Vernon}
\email{alexander.j.vernon@kcl.ac.uk}
\affiliation{Department of Physics, King's College London, Strand, London WC2R 2LS, UK}
\affiliation{London Centre for Nanotechnology}

\author{Sebastian Golat}
\affiliation{Department of Physics, King's College London, Strand, London WC2R 2LS, UK}
\affiliation{London Centre for Nanotechnology}

\author{Francisco~J. Rodr\'iguez-Fortu\~no}
\email{francisco.rodriguez\_fortuno@kcl.ac.uk}
\affiliation{Department of Physics, King's College London, Strand, London WC2R 2LS, UK}
\affiliation{London Centre for Nanotechnology}

\maketitle

This short supplementary document characterises the polarisation ellipse drawn by each of the $\mathbb{C}^3$ fields of the three natural bispinor bases presented in the main article, calculating the ellipse normal vectors and rectifying phases for each.

The three natural bases in which to represent the electromagnetic bispinor, $\bisp$, are the (traditional) electric/magnetic (EM) basis \cite{Golat2024},
\begin{equation}\label{EM_bispinor}
    \bisp_{\text{EM}}(\mathbf{r})=\begin{pmatrix}\mathbf{F}_\text{e}\\\mathbf{F}_\text{m}\end{pmatrix}=\frac{1}{2}\begin{pmatrix}\sqrt{\varepsilon_0}\mathbf{E}\\\sqrt{\mu_0}\mathbf{H}\end{pmatrix},
\end{equation}
the right-/left-handed basis,
\begin{equation}\label{LR_bispinor}
    \bisp_\text{RL}(\mathbf{r})=\pmqty{\mathbf{F}_\text{R}\\\mathbf{F}_\text{L}}=\frac{1}{2\sqrt{2}}\pmqty{\sqrt{\varepsilon_0}\mathbf{E}+i\sqrt{\mu_0}\mathbf{H}\\\sqrt{\varepsilon_0}\mathbf{E}-i\sqrt{\mu_0}\mathbf{H}},
\end{equation}
and the parallel/anti-parallel basis,
\begin{equation}\label{PA_bispinor}
    \bisp_\text{PA}(\mathbf{r})=\pmqty{\mathbf{F}_\text{p}\\\mathbf{F}_\text{a}}=\frac{1}{2\sqrt{2}}\pmqty{\sqrt{\varepsilon_0}\mathbf{E}+\sqrt{\mu_0}\mathbf{H}\\\sqrt{\varepsilon_0}\mathbf{E}-\sqrt{\mu_0}\mathbf{H}}.
\end{equation}
By virtue of their being phasors any vector-valued coefficient $\mathbf{F}_i(\mathbf{r})=\mathbf{a}_i(\mathbf{r})+i\mathbf{b}_i(\mathbf{r})$ of a component of the bispinor $\bisp$, whatever the basis and where $\mathbf{a}_i$ and $\mathbf{b}_i$ are real, inherits the same well-known, generic local geometry of the instantaneous electric and magnetic field vectors, which is to say that the instantaneous vector,
\begin{equation}
    \inst{F}_i(\mathbf{r},t)=\Re\{\mathbf{F}_i(\mathbf{r})\exp(-i\omega t)\}=\mathbf{a}_i(\mathbf{r})\cos\theta+\mathbf{b}_i(\mathbf{r})\sin\theta,
\end{equation}
usually draws an elliptical shape over one period.
Any such ellipse can be characterised by its normal vector,
\begin{equation}\label{normal}
    \mathbf{n}_i(\mathbf{r})=\mathbf{a}_i(\mathbf{r})\times\mathbf{b}_i(\mathbf{r})=\frac{1}{2}\Im\{\mathbf{F}^*_i\times\mathbf{F}_i\},
\end{equation}
and rectifying phase \cite{Berry2001},
\begin{equation}\label{rectifying_phase}
    \chi_i=\frac{1}{2}\arg{\phi_i}=\frac{1}{2}\arg{\mathbf{F}_i\cdot\mathbf{F}_i},
\end{equation}
two parameters that are well-defined provided $\mathbf{F}_i$ is not linearly polarised, where $\mathbf{n}_i=\mathbf{0}$ and its direction is undefined, or circularly polarised where $\mathbf{F}_i\cdot\mathbf{F}_i=0$.
The rectifying phase $\chi_i$, ill-defined when $\mathbf{F}_i\cdot\mathbf{F}_i=0$, is the phase angle for which the vectors $\mathbf{a}'_i$ and $\mathbf{b}'_i$ of $\mathbf{a}'_i+i\mathbf{b}'_i=\mathbf{F}_i\exp(-i\chi_i)$ are aligned to the ellipse semi axes.
When in the standard electric-magnetic basis, \eqref{normal} for $i=(\text{e, m})$ produces quite familiar expressions, 
\begin{equation}\label{normal_FEM}
\begin{split}
    \mathbf{n}_\text{e}=\frac{1}{8}\varepsilon_0\Im\{\mathbf{E}^*\times\mathbf{E}\},\\
    \mathbf{n}_\text{m}=\frac{1}{8}\mu_0\Im\{\mathbf{H}^*\times\mathbf{H}\},
\end{split}
\end{equation}
that combine with prefactors to give the electromagnetic field's total spin angular momentum density,
\begin{equation}\label{SAM}
    \mathbf{S}=\frac{1}{4\omega}\Im\{\varepsilon_0\mathbf{E}^*\times\mathbf{E}+\mu_0\mathbf{H}^*\times\mathbf{H}\}=\frac{2}{\omega}(\mathbf{n}_\text{e}+\mathbf{n}_\text{m}).
\end{equation}
Normal vectors and rectifying phases can be calculated for the polarisation ellipses of pure helicity fields and the parallel and anti-parallel fields by substituting the definitions of $\mathbf{F}_i$ for $i=(\text{R, L, p, a})$ from \eqref{LR_bispinor} and \eqref{PA_bispinor} into \eqref{normal} and \eqref{rectifying_phase}.
Doing so gives four equations,
\begin{equation}\label{normal_FRL}
    \mathbf{n}_\text{R/L}=\frac{1}{4}\left(\omega\mathbf{S}\pm\frac{1}{c}\Re\{\mathbf{\Pi}\}\right),
\end{equation}
\begin{equation}
    \mathbf{n}_\text{p/a}=\frac{1}{4}\left(\omega\mathbf{S}\pm\frac{1}{c}\Im\{\mathbf{\Pi}\}\right),
\end{equation}
\begin{equation}\label{phi_RL}
    \phi_\text{R/L}=\frac{1}{8}\left(\varepsilon_0\mathbf{E}\cdot\mathbf{E}-\mu_0\mathbf{H}\cdot\mathbf{H}\pm i\frac{2}{c}\mathbf{E}\cdot\mathbf{H}\right),
\end{equation}
\begin{equation}
    \phi_\text{p/a}=\frac{1}{8}\left(\varepsilon_0\mathbf{E}\cdot\mathbf{E}+\mu_0\mathbf{H}\cdot\mathbf{H}\pm \frac{2}{c}\mathbf{E}\cdot\mathbf{H}\right),
\end{equation}
which differ between the two fields of a certain basis by the $\pm$ sign of the final term, and which contain relations not only to the time-averaged total spin angular momentum density \eqref{SAM}, but to the real and imaginary \cite{Xu2019,Zhou2022} parts of the complex time-averaged Poynting vector,
\begin{equation}
    \mathbf{\Pi}=\frac{1}{2}\mathbf{E}^*\times\mathbf{H}.
\end{equation}
Through general arguments of how many of the total spatial degrees of freedom (two total in paraxial light, three in non-paraxial light) need to be spent satisfying $\mathbf{E}\cdot\mathbf{E}\propto\mathbf{F}_\text{e}\cdot\mathbf{F}_\text{e}=0$ and $\mathbf{n}_\text{e}=\mathbf{0}$, it is understood that electric circular polarisation normally occurs at confined points in the transverse plane of paraxial light (C points) and along lines in non-paraxial light (C lines), while linear polarisation in $\mathbf{E}$ is identified along lines (L lines) in both regimes where $\Re\{E_x\}/\Im\{E_x\}=\Re\{E_y\}/\Im\{E_y\}=\Re\{E_z\}/\Im\{E_z\}$ is satisfied (one of these conditions is relaxed in paraxial light as polarisation becomes 2D).
While those same arguments could be applied in principle to any basis coefficient field $\mathbf{F}_i$, the fields $\mathbf{F}_\text{R/L}$ are so intimately tied to symmetries of the electromagnetic field that are held in far fields, yet broken in the non-paraxial regime, as to depart from the familiar paraxial polarisation structure of the electric and magnetic fields.

To see why, we identify the three terms of $\phi_\text{R/L}$ \eqref{phi_RL}, which amount to zero when $\mathbf{F}_\text{R/L}$ is circularly polarised, with oscillatory parts of dot products of instantaneous field vectors.
The first two terms of \eqref{phi_RL} relate to instantaneous electric $\mathcal{W}_\text{e}$ and magnetic energy density $\mathcal{W}_\text{m}$, for example,
\begin{equation}
    \mathcal{W}_\text{e}(\mathbf{r}, t)=\frac{1}{2}\varepsilon_0\inst{E}\cdot\inst{E}=\frac{1}{4}\varepsilon_0\mathbf{E}^*\cdot\mathbf{E}+\frac{1}{4}\varepsilon_0\left(\Re\{\mathbf{E}\cdot\mathbf{E}\}\cos{2\omega t}+\Im\{\mathbf{E}\cdot\mathbf{E}\}\sin{2\omega t}\right),
\end{equation}
\begin{equation}
    \mathcal{W}_\text{m}(\mathbf{r}, t)=\frac{1}{2}\mu_0\inst{H}\cdot\inst{H}=\frac{1}{4}\mu_0\mathbf{H}^*\cdot\mathbf{H}+\frac{1}{4}\mu_0\left(\Re\{\mathbf{H}\cdot\mathbf{H}\}\cos{2\omega t}+\Im\{\mathbf{H}\cdot\mathbf{H}\}\sin{2\omega t}\right),
\end{equation}
while the third term of \eqref{phi_RL}, $\propto\mathbf{E}\cdot\mathbf{H}$, informs the time-dependence of the dot product of the instantaneous electric and magnetic vectors,
\begin{equation}\label{EdotH_instant}
    [\inst{E}\cdot\inst{H}](\mathbf{r},t)=\frac{1}{2}\Re\{\mathbf{E}^*\cdot\mathbf{H}\}+\frac{1}{2}\left(\Re\{\mathbf{E}\cdot\mathbf{H}\}\cos{2\omega t}+\Im\{\mathbf{E}\cdot\mathbf{H}\}\sin{2\omega t}\right).
\end{equation}
For as long as there are equal double-frequency oscillatory components to the instantaneous electric and magnetic energy density (implying $\varepsilon_0\mathbf{E}\cdot\mathbf{E}=\mu_0\mathbf{H}\cdot\mathbf{H}$) and the instantaneous vectors $\inst{E}(\mathbf{r},t)$ and $\inst{H}(\mathbf{r},t)$ are always at right angles (implying \eqref{EdotH_instant} $\inst{E}\cdot\inst{H}=0$), then the complex scalar fields $\phi_\text{R/L}$ are zero.
When $\phi_\text{R/L}=0$, the rectifying phases $\chi_\text{R/L}$ from \eqref{rectifying_phase} are undefined indicating that $\mathbf{F}_\text{R}$ and $\mathbf{F}_\text{L}$ are circularly polarised. 
All paraxial light, which satisfies both $\varepsilon_0\mathbf{E}\cdot\mathbf{E}=\mu_0\mathbf{H}\cdot\mathbf{H}$ and $\inst{E}\cdot\inst{H}=0$ regardless of polarisation in $\mathbf{E}$ and $\mathbf{H}$, must always be wholly circularly polarised in the pure helicity states $\mathbf{F}_\text{R}$ and $\mathbf{F}_\text{L}$---in \eqref{normal_FRL}, even if the $\mathbf{E}$ and $\mathbf{H}$ fields are linearly polarised and reduce SAM density $\mathbf{S}$ to zero, both normal vectors $\mathbf{n}_\text{R/L}$ remain non-zero because of the Poynting vector $\mathbf{P}$.

When in the non-paraxial regime the electric and magnetic fields gain three components, the spatial structure of polarisation of all six basis coefficient fields in Eqs.~(\ref{EM_bispinor})-(\ref{PA_bispinor}) complicates and contains one-dimensional polarisation singularities, analogies of C lines and L lines, including even the pure helicity fields $\mathbf{F}_\text{R}$ and $\mathbf{F}_\text{L}$ that can, if counter-intuitively, become linearly polarised.
